\documentclass[twocolumn]{aastex62}
\usepackage{color} 
\usepackage{lipsum}
\usepackage{mathtools}
\usepackage{placeins}
\usepackage{multirow}
\usepackage{float}
\usepackage{amsmath}
\usepackage{physics}
\usepackage{silence}
\usepackage{multirow, booktabs}
\usepackage{endnotes}

\shorttitle{The First Global e-Competition on Astronomy and Astrophysics}

\begin{document}
\title{The first Global e-Competition on Astronomy and Astrophysics}


 \author{Ioana A. Zelko}
 \thanks{corresponding author, ioana.zelko@gmail.com}
 \affiliation{Canadian Institute for Theoretical Astrophysics, University
of Toronto, Toronto, Canada}
\affiliation{Department of Physics and Astronomy, University of California-Los Angeles, Los Angeles, California, United States of America}
 \affiliation{Harvard-Smithsonian Center for Astrophysics, 
 Cambridge, Massachusetts,  
 United States of America}
 
 \author{Charles Barclay}
 \affiliation{Marlborough College, Marlborough, United Kingdom}

     \author{Tõnis Eenmäe}
 \affiliation{Tartu Observatory, University of Tartu,  Estonia}
 \author{Taavet Kalda}
 \affiliation{University of Oxford, Oxford, United Kingdom}
 \author{Hara Papathanassiou}
 \affiliation{Leiden Institute of Physics, Leiden University, 
 Leiden,  The Netherlands}
 \author{Nikita Poljakov}
 \affiliation{University of Bristol, Bristol, United Kingdom}
 \author{Gustavo A. Rojas}
 \affiliation{NUCLIO - Núcleo Interactivo de Astronomia - São Domingos de Rana, Portugal}
 \affiliation{Núcleo de Formação de Professores - Universidade Federal de São Carlos - São Carlos, Brazil}
 \author{Tiit Sepp}
 \affiliation{University of Tartu, Tartu, Estonia}
 \affiliation{STACC, Tartu, Estonia}
 \author{Greg Stachowski}
 \affiliation{Cracow Pedagogical University, Krakow, Poland}
 \author{Aniket Sule}
 \affiliation{Homi Bhabha Centre for Science Education, Mumbai, India}

\begin{abstract}
The first Global e-Competition on Astronomy and Astrophysics was held online in September-October 2020 as a replacement for the International Olympiad on Astronomy and Astrophysics, which was postponed due to the COVID-19 pandemic. Despite the short time available for organisation, 8 weeks, the competition was run successfully, with 325 students from over 42 countries participating with no major issues. The feedback from the participants was positive and reflects the ways in which such events can boost interest in astronomy and astronomy education. With online activities set to become more prevalent in the future, we present an overview of the competition process, the challenges faced, and some of the lessons learned, aiming to contribute to the development of best practices for organizing online competitions. 

\end{abstract}


\section{Introduction}

\begin{figure}[t!]
\includegraphics[width=\columnwidth]{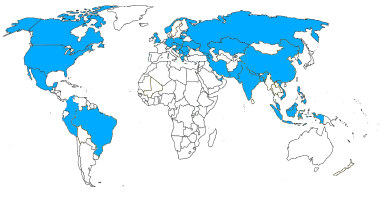}
\caption{Countries Participating in the 1$^{\rm st}$ GeCAA. The event attracted 325 students from 42 countries.
}\label{fig:worldmap}
\end{figure}

The world of physics education faces significant challenges in maintaining students' interest and fostering a sense of self-efficacy among budding scientists. Numerous studies have highlighted the negative impact of introductory physics courses on students' perceptions of physics and their view of themselves as scientists \cite{ Redish1998, Adams2006,Taibu2020}. Moreover, these studies also report a tendency for student expectations to deteriorate rather than improve as a result of the first term of introductory calculus-based physics.\\
To counter these challenges and promote interest in the field, scientific competitions have long been recognized as valuable tools for engaging students in various scientific disciplines \cite{burguillo_using_2010, cantador_effects_2010,oliver_exploratory_2011,wirt_analysis_2011,sukiman_competition-based_2016}. These competitions offer students opportunities to challenge themselves, broaden their horizons, and connect with like-minded individuals from around the world.\\
In this paper, we present the results of the inaugural Global e-Competition on Astronomy and Astrophysics, which took place online in September-October 2020 as a replacement for the International Olympiad on Astronomy and Astrophysics. A follow-up survey showed that the competition had a quantified positive impact on participants' views of astronomy and astrophysics. This centers the competition as an important tool for attracting students to this career path. As such, it serves as an excellent case study for physics educators looking to engage and inspire students in their respective disciplines. \\
Our analysis of the competition's global participation and narrative addresses another critical topic in education: diversity, equity, and inclusion. The event attracted 325 students from over 42 countries (see Figure \ref{fig:worldmap}), contributing to the development of a thriving scientific community and encouraging future generations of researchers, educators, and professionals.\\
The tasks and challenges presented at GeCAA often reflect the cutting edge of scientific inquiry in data analysis problems and group projects. Previous research has shown that students' incoming preparation in physics correlates well with their performance in introductory undergraduate classes \citep{Salehi2019}. Thus, the competition can serve as effective preparation for college students. Furthermore, the competition provides college professors a wealth of resources for staying current with the latest developments in astronomy and astrophysics and for incorporating new concepts and methods into their courses. By analyzing students' performance on these tasks, professors can also gain insights into the efficacy of their teaching methods and identify areas where improvements may be made.\\
In the following sections, we will discuss the competition process, tasks, challenges faced, and some of the lessons learned, aiming to contribute to the development of best practices for organizing online competitions and bringing awareness to the resources provided by the competition.

\section{Design Principles}

The International Olympiad on Astronomy and Astrophysics \citep[IOAA,][]{1ioaa,Soonthornthum2011,2019EPJWC.20001011S} is an annual event during which senior secondary-school students compete, solving theoretical and practical problems in astronomy. It was established in response to a perceived need for an event similar to the International Physics Olympiad (IPhO) but more specifically tailored to astronomy and astrophysics \citep{1ioaa}. As such, it forms one of the International Science Olympiads.

At IOAA, any country may send a team of up to five students, accompanied by two adult team leaders. With the host's permission, a second or "guest" team may also be sent. 
The host country organizes the competition and provides accommodation and lectures, cultural events and tours for the participants. These events
encourage bonding among like-minded youngsters and create an international community of young astronomy enthusiasts. This aspect has been consistently highly valued by participants, many of whom go on to study and work in astronomy following this positive experience. The IOAA has also been the motivator for establishing national astronomy competitions  in several countries, with  IOAA past exams serving as resources for instructors. Lastly, the challenges of hosting the IOAA have led to closer ties between professional and amateur astronomers, as well as between astronomers and teachers. All these aspects help to develop astronomy education in participating countries \citep{2019EPJWC.20001011S}.

Participating students are tasked with solving problems covering the theory of astronomy and astrophysics (50\% of the final score), practical data analysis using real or simulated data (25\%), and practical night-sky observation (25\%), including outdoor and planetarium-dome components. The tasks are set by the host country, approved by the International Board and then translated into the students’ native languages. Marking is performed by a local jury and verified through moderation with the team leaders. Gold, silver and bronze medals along with honourable mention certificates are awarded to students who achieve appropriate scores, and special prizes may additionally be awarded for top results in different categories. 

A separate component of the competition is team challenge, in which students are pseudo-randomly allocated into international teams of between five and eight members. The teams are then assigned a problem which requires collaboration to solve, thereby creating  additional opportunities for peer interaction and simulating the international nature of science in general and astronomy in particular. This aspect of the competition has consistently proven to be very popular with participants, and is unique to IOAA.

From the inaugural event in 2007 until 2019, the IOAA was successfully held every year. During this period, it has been hosted in Thailand, Indonesia, Iran, China, Poland, Brazil, Greece, Romania, India and Hungary (with China, Indonesia and Thailand hosting the IOAA twice). Participation has increased from 85 participants from 22 countries in 2007 to 259 participants from 46 countries in 2019. Additionally, 8 other countries have participated intermittently, with several more sending observers or expressing interest in future participation. 

In June 2020, it became apparent that the planned 14th IOAA competition in Bogotá, Colombia, would not take place due to the pandemic. After consultation with leaders of the national teams, the board of the IOAA decided to organize an online competition in order not to deprive the interested (and in many cases already selected) students the opportunity of competing internationally, and also provide continuity for the competition and maintain involvement of the community of volunteers.
A new name was therefore suggested to reflect the online nature of the competition: the Global electronic Competition on Astronomy and Astrophysics (GeCAA). 

Traditionally, the IOAA has not charged participation fees. 
GeCAA followed the same financial model, with no participation fee for the teams. Team leaders were asked to volunteer to help in the grading of the answer scripts.

Although the circumstances leading to this online event were unprecedented, the novel venture was successful and a substantial learning experience for all those involved. We report on it here to assist in the organization of similar events and to highlight potential pitfalls that can be avoided, thereby increasing the potential for success of any online competition, whether fully or partially remote.

While conceptualizing a new online competition, the organisers had to consider several constraints. As it was intended to replace the IOAA, the organizers had to take into account that participants (both students and mentors) would be familiar with the academic structure of the IOAA. For this reason, GeCAA was designed
to remain as close to the familiar IOAA structure as possible.

\subsection{Student Preparation}
To prepare for the competition,  students cover material that spans high school physics to college-level astronomy and astrophysics. Some of the resources used are listed at \url{https://usaaao.org/resources/} ). The levels of the problems in the competition are suitable for homework given in classes at the college level
. Participants cover the field broadly, as it would be in an introductory level astronomy course. However, they also train on many sets of problems that give them depth as well.
In addition to theory, a big part of the preparation is learning to observe the night sky, in a similar fashion to amateur astronomers. Students often seek to practice at college or amateur astronomy clubs, and become very proficient in this subject.  

\subsection{National Competitions}
In order to send students to the international competition, countries hold national competitions, with variations in structure between countries. As a general pattern, students participate in different levels of selections from local, to regional, to national training camps\endnote{
For reference, the national selection process for the USA is described on the USAAAO website at \url{https://usaaao.org/about/selection-process/}, and briefly summarized in the Supplemental Material in Section 4.}. 
The pandemic led to lock-downs in most countries and their national selection process was incomplete. In most cases, the countries had shortlisted 10-20 top students from their national competitions, but could not proceed with further filtering due to sudden lock-downs. At the same time, the online nature of GeCAA enabled the participation of more students from each country than the normal five-member teams without additional cost. Thus, it was decided to allow a maximum of 10 students per country for individual competition rounds and a maximum of 10 students for the team competition round. The countries were given the freedom to register different participants for the individual and the team competitions, effectively allowing up to 20 students from each country to benefit from participation and allowing for more flexibility around students' schedules.


\subsection{International Competition}

The schedule of the competition can be found 
at \url{https://gecaa.ee/timetable/}. The team competition started after the individual rounds finished over the first 3 days. The results were announced a couple of days after the team competition finished, as soon as the marking was done. The online lectures and cultural activities took place after the individual rounds were done, and before the ending ceremony.

\subsubsection{Individual Competition}
All three components (theory, data analysis, and knowledge of the night sky) of an IOAA had to be included. These tests were conducted over three consecutive days, including a weekend, in late September 2020. Similar to IOAA, the students had question papers in their own languages and the solutions to the questions were limited to mathematical expressions and calculations which could be assessed without significant translation. Some questions had to be significantly rephrased to be answerable as either numerical values, multiple choice, or a single word to maximise the use of automated marking deployed in the custom-made website for the competition. (This is described in the section called ``Implementation'' in the Supplementary Materials). These constraints guided development of the online examination tool described in the next section. Namely, the tool had to allow easy translations and a seamless switch between English and the native language. This was achieved by making the original English version of tasks available as a {\emph {Google}} document to the mentors,  who then submitted the translations in separate {\emph {Google}} documents with the same formatting. It was also clear that students would be most comfortable working on paper, so rather than using sophisticated input methods, the focus was on enabling students to scan and upload their work onto the server.

For any online international event, the biggest challenge is to provide equal opportunity to participants from different time zones. This is even more crucial for examinations or competitions where safeguarding academic integrity can be complicated. It is not possible to schedule simultaneous tests for all time-zones without causing significant inconvenience to some participants. In the case of GeCAA, the participating countries spanned from South Korea to Colombia. As a solution, tests were conducted with 5 different starting times between 08:00 UTC to 15:00 UTC each day. The responsibility of maintaining the academic integrity was placed with the national olympiad committees. The students were not allowed access to any informational resources, such as books or websites, during the exam. The national committees proctored the exams and recorded the rooms and the students with audio and video. These recordings were subsequently verified by the GeCAA organizers.

\begin{table*}[th!]
\renewcommand{\arraystretch}{1.0}
\setlength{\tabcolsep}{3pt}
\centering
\begin{tabular}{lll|lll}
\hline
\hline
\multicolumn{6}{c}{\textbf{Individual Competition}}  \\
\multicolumn{3}{c}{\textbf{Theory Problems}} & \multicolumn{3}{c}{\textbf{Data Analysis Problems}}\\
\# & Title  & Field &  \# & Title  & Field \\ 
1 & Astrophotography & Observations, Instruments & 1 & Active Galactic Nuclei (AGN)& Galaxies \\
2 & Flat Earth & Radiation Mechanisms & 2 & Minor Planet& Exoplanets \\
3 & Mirror & Cosmology & 3 & Hypervelocity stars & Stellar Science \\
4 & Light Curves & Binary Stars & \multicolumn{3}{c}{\textbf{Observation Problems}}\\
5 & HII Region & The Interstellar Medium &1& Comets in the “air” & Stars, Constellations\\
6 & Occultation of a X-ray Source & Positional Astronomy & 2 & Neowise with MAGIC & Comets \\
7 & Radiant of a Meteor Shower & Spherical Astronomy & 3 & All Sky & Celestial Sphere\\
8 & Jupiter's Great Red Spot & Planetary Science & 4 &  Sky Map & Celestial Sphere  \\
\hline
\multicolumn{6}{c}{\textbf{Team Competition}}\\
 \multicolumn{2}{r}{\#} & \multicolumn{2}{l}{Title} & \multicolumn{2}{l}{Field}\\
 \multicolumn{2}{r}{1} & \multicolumn{2}{l}{Spectral Distortions of the CMB } & \multicolumn{2}{l}{Cosmology}\\
  \multicolumn{2}{r}{2} & \multicolumn{2}{l}{Measuring the Distance to the Moon} &\multicolumn{2}{l}{ Parallax Observations} \\
\hline
\end{tabular}
\caption{The problems presented to the students at GeCAA during the 4 academic tasks of the competition: theory, data analysis, observation, and team competition. The tasks range in difficulty from typical introductory college coursework to graduate coursework. Both the tasks and the solutions can be found at the competition's website at \url{https://gecaa.ee/competition-problems-and-solutions/}.\label{table:problems}}
\end{table*}
Equal access to the internet also turned out to be a critical hurdle. Some of the participating countries had restrictions on the use of {\emph {Google}} products that were used to share the competition tasks. In such cases, the questions were sent to the national Olympiad committees by email and student answer sheets were also accepted by email. For many countries, even domestic travel was restricted and it was not possible to bring all student participants to the same venue. In such cases, students were allowed to participate in the examination from their homes. All participants were remotely proctored through Zoom calls with proctors manually monitoring the feeds. Zoom calls were recorded with the students' knowledge for later verification, if necessary. In case a country’s internet policy did not allow Zoom calls, the national Olympiad committees of such countries were asked to proctor their students with appropriate domestic software and submit the recording later. These measures covered most cases, except for one unexpected situation of a total internet shutdown in one of the participating countries, due to the sudden eruption of a border war.


\subsubsection{The Team Competition}


The team competition for GeCAA was conducted by 
43 teams (named after IAU constellations) of 6-7 students each. The participating countries were divided into 6 groups, relating to their geographic proximity and cultural similarity and each team included at least one student from each group. This also automatically meant that each team had students from a wide range of time zones (from East Asia to the Americas), which was crucial for one of the academic tasks for the team competition. The competition consisted of two tasks (described in Section \ref{sec:AcaTasks}), in which the teams needed to deliberate and submit a solution or a report after two weeks. The nature of the open-ended tasks eliminated the need for direct or indirect proctoring. Students were allowed access to any resources they could find. Moreover, the extended duration of the task allowed for enough opportunities for meaningful collaboration and sharing of work.

\subsubsection{Online Lectures}

The Estonian organizers  scheduled two lectures by an eminent astronomer and an astronaut for all the participants, which proved very popular among the students. Supplementing the program with lectures from international professionals, who will be seen as role models, is of immense value. Enriched with personal experience and advice, these talks covered various aspects of astronomy. Such lectures reinforce the educational character of the activity, provide unique opportunities to expose students to international experts and enable students' interaction with the speakers.


\section{Academic Tasks}
\label{sec:AcaTasks}

The academic tasks focused on four components: theory, data analysis, and observation for individual competition, plus the team competition (following the structure of IOAA). Table \ref{table:problems} presents the title and field of each problem in the tasks. The tasks range in difficulty from typical introductory college coursework to graduate coursework.

The problems for the individual exams were designed to be solved in the span of hours (2-4 hours per exam). Some examples of questions in the theoretical part follow: In question 4, contestants are given a schematic lightcurve of a fictitious eclipsing binary system and are asked to deduce the ratios of the values of stellar properties. In a series of modified lightcurves, they have to identify which of a given set of scenarios is responsible for each modification.

Question 8 of the theory exam looks at Jupiter’s Great Red Spot: Relying on a velocity map, the contestants are guided in deducing the shape, area, and vorticity of the spot and identify it as cyclonic. They then find the minimum displacement in latitude that will convert the Spot into an anticyclonic system. This requires the iterative solution of a nonlinear equation.

Problems in Data Analysis draw directly from recent research literature. Two examples are summarized here: Problem 1 gives a schematic of an active galactic nucleus and guides the contestants to deduce the distance of the Broad Line Region from the central black hole, using the light curves of the B continuum and the optical line spectrum. From the line spectrum, they obtain the dispersion velocity and calculate the mass of the central black hole with the use of the virial theorem.

In Problem 3, contestants first have to classify the spectrum of a hypervelocity star by comparison with template spectra. Then, given its luminosity class and galactic coordinates, they determine the distance to the star and its distance from the galactic center. From the spectral shift and proper motion, they obtain its velocity. They compare this to the escape velocity for a mass distribution of the halo, to conclude that the star is unbound and that, furthermore, it has travelled for a time comparable to its age.

In contrast to the short timeline of the individual competition, students were given two weeks to solve the two tasks of the team competition. 
Problems in the group competition rely on the collaboration of its members. The first problem encouraged students to explore the scientific literature and become familiar with topics beyond the standard curriculum in order to understand what information was required to answer the questions. The first topic was that of spectral distortions of the cosmic microwave background (CMB) \citep[e.g.][and references therein]{Zeldovich1969,Sunyaev1972, Tashiro2014}, which are departures from the black body spectrum of the CMB. Students were asked to investigate the possibility of the proposed satellite Primordial Inflation Explorer \citep[PIXIE]{Kogut2011, Kogut2016, Naess2019,Kogut2020} obtaining a detection  of the primordial spectral distortions \citep{Abitbol2017, Zelko2021}. The second problem required students to take advantage of their different positions across the globe and combine night sky measurements from all of their geographical locations to obtain a solution to the problem. They were able to experimentally determine the distance to the Moon using parallax measurements, as a result of their combined efforts.
The teams had to devise, execute, and document an observing program, and present their results in a short report.


\begin{figure}[t!]  
\includegraphics[width=\columnwidth]{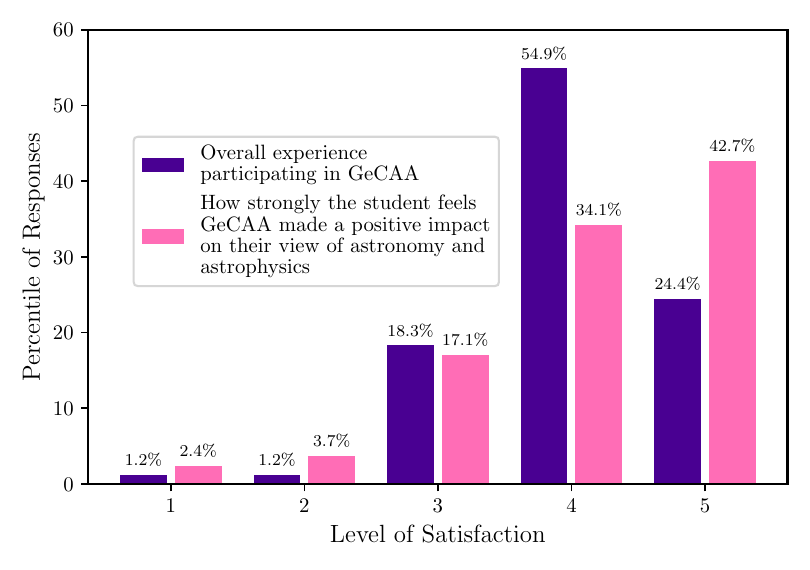}
\caption{Student reviews for the overall experience participating in GeCAA, and the impact of GeCAA on their view of astronomy and astrophysics, are presented in the figure. The heavily positively-biased distributions confirm that GeCAA has succeeded in its primary goals. Rsponses were collected from 82 survey participants. 
}\label{fig:gecaa_rank}
\end{figure}
\section{Participation}
\label{sec:Participation}

Despite pandemic-induced local lockdowns, GeCAA attracted interest from 42 countries (see Figure \ref{fig:worldmap}). A total of 278 students participated in the individual rounds of GeCAA and 293 students took part in the team competition. India and Qatar exclusively participated in the team competition and several countries extended the opportunity of participation to more students by nominating different students for individual and team competitions.  In total, 325 students were able to partake in the GeCAA experience.

\section{Lessons Learned and Follow-up}
\label{sec:Lessons}
\paragraph{Lessons Learned}
We conducted a survey of the participants (see Fig. \ref{fig:gecaa_rank}), the results of which indicate that most students found the event to have been a great experience. The event positively influenced their view of astrophysics, confirming that the goals of GeCAA were achieved. The full details of the survey results can be seen in the Supplementary Material. The survey suggests that the competition plays a pivotal role in drawing students towards this career path. It stands as a valuable example for physics educators aiming to captivate and motivate students. \\ 
The feedback provided includes lessons that could serve the community of physics teachers as a whole. For future events, students recommended incorporating programming and software-based problems. This insight can be useful to university-level teachers, as current cutting-edge physics research in many fields relies heavily on programming. Students also requested extending the theoretical exam (closer to the original five-hour theoretical exam in IOAA), which indicates the appeal of problems that can take longer to solve and have depth. They also suggested offering exams in PDF format (as opposed to the competition website) for quicker problem solving. Additional suggestions can be found in the Supplemental Materials.\\
While the event was successful, certain aspects could have been improved.  The following points are pertinent to any teacher who is involved in creating exams:  have individuals solve each problem before finalization, to catch potential mistakes; test tasks involving approximations or curve-fitting for numerical stability;  before grading begins, hold a mandatory meeting between the graders and the problem writers for each question, to synchronize the grading scheme and reduce bias; supplement the marking scheme for each question with a robust way of carrying an erroneous numerical result through the calculation via a spreadsheet or Jupyter Notebook (in order to avoid repeatedly penalizing early mistakes).\\
Finally, for competition organizers, in addition to the points made above, there are more useful lessons. Redundancy should be included in the planning staff assuming some degree of grader attrition. In addition, this competition  was organised on short notice of 8 weeks, and such a heavy burden on a few people is not a sustainable model. Organizers should plan events with a longer timeline and communicate expected time commitments clearly to all volunteers.
\paragraph{Follow-up}
In recent years, traditional academic Olympiads, including the IOAA, have shifted towards embracing online and hybrid formats. Before 2020, most of these competitions were averse to remote participation. However, the various models of online participation piloted in the 2020 GeCAA competition have since acted as guiding lights for the evolution of IOAA events. The 2021 Online IOAA saw teams participating remotely while grading was exclusively conducted by the host country. However, the most significant impact of the GeCAA model was observed during IOAA 2022. Initially scheduled to take place in Ukraine, the event was moved to Georgia due to geopolitical concerns. While 38 countries sent teams to participate in person, six teams could only join online. The examination schedule for these remote teams was staggered according to their respective time zones, following the principles established by the GeCAA. Furthermore, just as in the GeCAA, team leaders from these countries exchanged answer scripts among themselves for grading, alleviating the burden on the host country. The entire process was completed fairly and without any issues.
Based on these experiences, the IOAA International Board has now approved new rules that allow for online participation as a regular feature in all future editions of IOAA. This decision underscores the successful adoption of innovations from online competitions into on-site events, accommodating the changing needs and preferences of participants and organizers alike.


In conclusion, the GeCAA's flexible format, achieved without compromising the educational character of a traditional Olympiad, holds value in a world where many activities are transitioning to online and hybrid formats.  This perspective is shared by several members of the IOAA executive committee and international board of team leaders, although it may not necessarily represent the views of every individual in the group. Furthermore, the experience gained from organizing GeCAA can inform the organization of other online competitions, irrespective of their scale.



\paragraph{Availability of supporting data and materials}
The academic tasks of the GeCAA are freely available on the GeCAA website\footnote{\url{https://gecaa.ee}}. More information about IOAA can be found on its website\footnote{\url{ https://www.ioaastrophysics.org}}. Resources on training for this competition can be found at the USA Astronomy and Astrophysics Olympiad website \footnote{\url{ https://usaaao.org/resources/}}.


\paragraph{Competing Interests} The authors declare that they have no competing interests.

\paragraph{Funding} Funding for organising this event in Estonia was provided by the Republic of Estonia Ministry of Education and Research, the Tartu Observatory of the University of Tartu, the Jaan Tallinn foundation, and the Estonian Research Council.

\paragraph{Author's Contributions}
All of the authors were either the AC or LAC members for the GeCAA event and set and assessed tasks during the competition. The following authors were primarily responsible for particular sections of the paper:  Ioana Zelko created and analysied the student survey, contributed to the Academic Tasks and Evaluation Process section, researched the references, and handled the submission of the paper and extended revision process. Hara Papathanassiou wrote most of the Advice for Future events and contributed to the Academic Tasks and Evaluation Process section. Nikita Poljakov was responsible for the Local Academic Committee and Evaluation section. Tiit Sepp wrote the description of the technical part of the event and Tõnis Eenmäe contributed to the technical description part. Greg Stachowski wrote the Introduction and Abstract. Aniket Sule wrote the Design Principles, and the Follow-up. All authors worked on editing and polishing of different sections and participated in deliberations during the drafting of the manuscripts.

\paragraph{Acknowledgements}

We extend our deepest appreciation to all members of the Academic Committee (AC) including Siramas Komonjinda, Mikhail Kuznetsov, Volodymyr Reshetnyk, Geraldine Tan, and Sonja Vidojevic. Further recognition is due to the Local Academic Committee (LAC) led by T. Kalda, N. Poljakov, T. Eenmäe, and J. Laur, along with the University of Tartu Youth Academy. Their collective effort in organizing the inaugural GeCAA, assistance in preparing this paper, and dissemination of the survey was invaluable. Our gratitude extends to Grete Lilijane Küppas, Gutnar Leede, and Mihkel Kree for their enlightening input on the technical facets of the engine used in the GeCAA competition, reflecting their integral roles as developers. We acknowledge the contribution of P. Kerner and R. Muser for their keen eye and support in the enhancement of the visual elements of the 1st GeCAA. Moreover, the success of the event owes much to the tireless work of countless volunteer graders, task proposers, and national team leaders, whose efforts we sincerely appreciate. Finally, we express our thanks to editors Prof. Beth Parks and Prof. Jesse Kinder, referee Prof. Dr. Vivek Narayanan, and two anonymous referees. Their insightful suggestions were instrumental in improving this article.

\bibliography{paper-refs.bib}



\begin{center}
    \Large
    \textbf{Supplementary Material}\\
\end{center}

\section*{Survey of the Participants}
\label{sec:Survey}


\begin{figure*}[t!]  
\hspace{-5mm}
\includegraphics[scale=1]{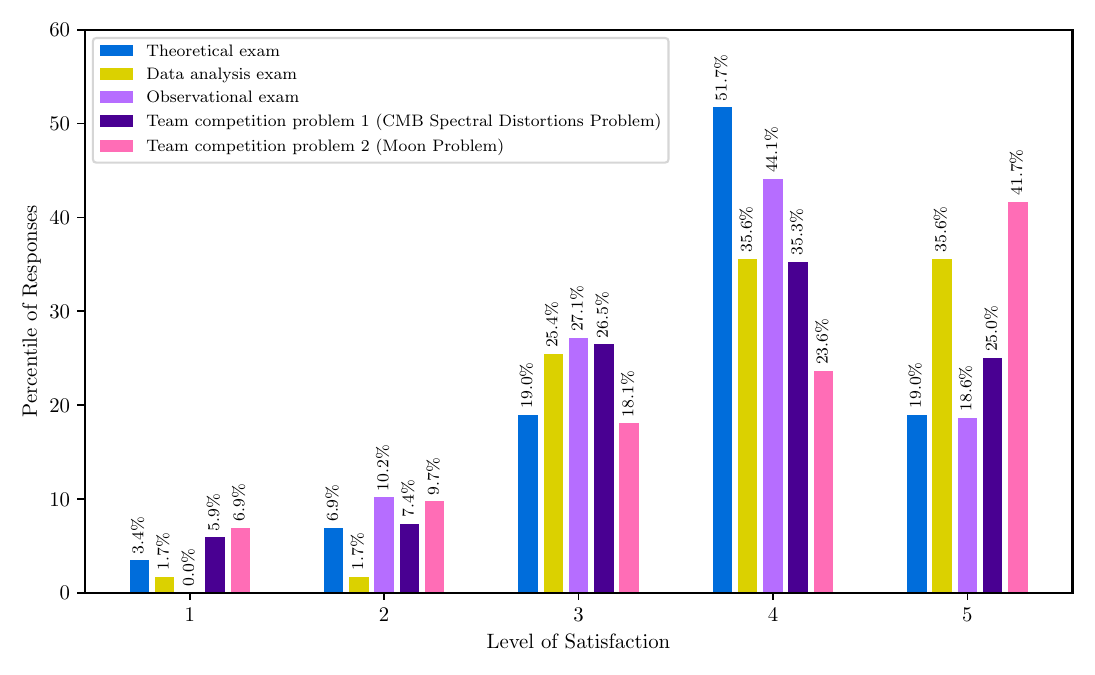}
\caption{Students' reviews of their experience for the 4 different exams: theoretical, data analysis, observational, and team competition.
}\label{fig:gecaa_exams_rank}
\end{figure*}

\begin{figure*}[t!]  
\hspace{-5mm}
\includegraphics[scale=1]{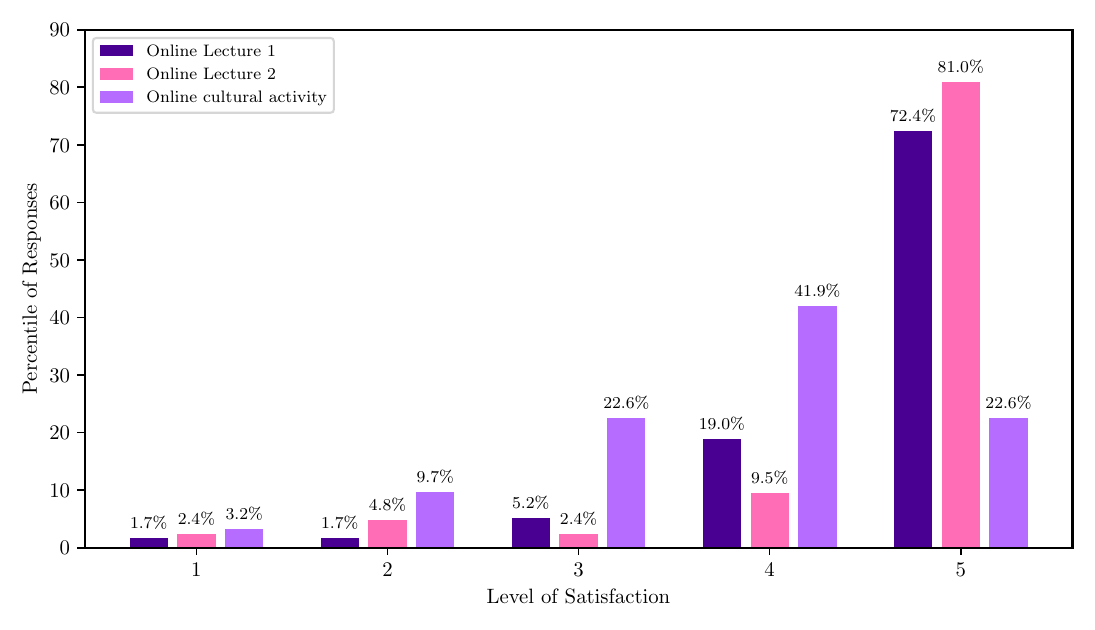}
\caption{The student reviews for the online Lectures and cultural activity events. The survey shows a very strong positive response to the two online lectures which students took part in; the lectures received the highest percentile of maximum level of satisfaction amongst all of the events of the competition. 
}\label{fig:gecaa_events_rank}
\end{figure*}
We conducted a survey of the participants a few months after the competition ended. The aim of the survey was to gauge the impact of the competition on the students, as well as to get feedback for possible future editions. The survey contained 27 questions, spanning the topics of general GeCAA participation experience, impact on the students’ interest in astronomy and astrophysics, online experience, exam preparation strategy, individual exam tasks, team competition, online lectures, cultural activities, and recommendations for the future. The questions targeted both long text answers and multiple choice rankings. 

In total, 82 student participants responded to the survey. The overall picture reflected in the answers was that the students had a positive experience and that the goals of GeCAA were achieved. For example, to the question `Please rate your overall experience participating in GeCAA' on a scale from 1 to 5 spanning “poor” to “great”, 97.6\% of survey participants gave a ranking greater than or equal to 3, and for the question `Please rate how strongly you feel GeCAA made a positive impact on your view of astronomy and astrophysics', 93.9\% of respondents gave a ranking greater than or equal to 3, with 42.7\% choosing the maximum ranking of 5, as can be seen in Fig. 3 of the main paper. 
Figures  \ref{fig:gecaa_exams_rank} and \ref{fig:gecaa_events_rank} show the quantitative reviews students gave to the exams and events they participated in, which indicate an overall positive response to all of the events.

Qualitatively, the students described GeCAA experience with phrases like:
\begin{itemize}
\item feeling like a scientist
\item getting to know interesting peers from other countries
\item learning from the online guest lecturers
\item learning from each other
\item solving fun challenges
\item learning the value of teamwork
\item having an overall fun and rewarding experience.
\end{itemize}

Students’ expectations from future events included:
\begin{itemize}
\item more recognition for individual performance in the team competition
\item more emphasis be put on each individual’s contribution to the teamwork in the team competition
\item online lectures and study materials that can be used for the year-long preparation be made available online
\item more social events be organized for participants' bonding
\item more online lecture webinars by researchers
\item more guidance for strategies and materials students should use to learn
\end{itemize}

\section*{Academic tasks and the evaluation process}

\subsection{The International Academic Committee}
The 12-member working group, formed to conceptualise the GeCAA, was maintained as the International Academic Committee (AC) for the 1st GeCAA. The names of the AC members were communicated to the International Board for their approval and the AC members were requested to recuse themselves from the training and selection of their respective national teams. The AC was tasked with the collection, grading and selection of the problems to be assigned. Suitable questions were sourced through voluntary contributions by various contributors from the national teams. Representatives from 16 countries submitted entries, many of which were multiple entries, and a total of 56 problems/questions were collected. These were each evaluated by at least two members of the AC and discussed during  a series of weekly online meetings. Of the 56 problems, 31 were categorised as theoretical, 7 as observational, 13 pertaining to data analysis and 5 were found to be suitable as team tasks. The questions spanned the following knowledge areas: celestial (planar) geometry, coordinate systems and spherical trigonometry, celestial mechanics, black body radiation, optical instruments, optical observations (magnitudes, photometry etc.), stellar spectroscopy, the Sun, planetary astrophysics, stellar physics, exoplanets, active galactive nuclei, and cosmology.

The AC convened online for eight sessions of 2-hours each in July -September with the following main topics:
\begin{enumerate}
\item Format and name of the competition
\item Scheduling of the academic programme 
\item Vetting of submitted questions
\item Short-listing of questions for the competition and assigning of each short-listed task to a reviewer (member of AC charged with independently solving the problem, editing the question, and working out a detailed marking scheme).
\item Discussion of observational questions and their implementation
\item Discussion of theoretical questions and their implementation
\item Discussion of data analysis questions and their implementation
\item Discussion of team tasks for the Team Competition. 
\end{enumerate}

The final tasks were typeset in the standard IOAA format (using a template in {\emph{Google Docs}}) and marking schemes in the IOAA format (LaTeX template). Most were checked by the author and the AC reviewer. Some of the tasks required substantial modification for clarity, level of difficulty and for making the solutions language-independent. The marking schemes were required to be rewritten in a detailed manner for all tasks. Solutions that accommodated for potential earlier mistakes were worked out by the reviewers and recorded in a spreadsheet or Jupyter Notebooks.

During the grading, several AC members were assigned the role of head examiner for different tasks and they contributed to resolving grading discrepancies alongside head examiners from the local academic committee.

\subsection{Local Academic Committee and Evaluation}
The Estonian national Olympiad committee formed a local academic committee (LAC) which was responsible for organising the grading process. Each member of the LAC was assigned to a single competition round. The grading process included three distinct tasks: 1) aiding graders during the marking; 2) resolving marking disagreements between the graders; 3) addressing student appeals.

Graders were invited from a voluntary pool of international team leaders (in contrast with the IOAA, where the organizing country provides the graders). They were organised into pairs responsible for grading certain competition problems following the marking scheme. The mapping of graders to the problems was done manually by taking into account the comfort level of the graders with different astrophysical topics and their prior experience with IOAA. Due care was given to ensure that the graders were not assigned answer scripts of their own country and two graders of each question were from disparate education systems to accommodate sensitivity to different styles of students’ solutions. At the start of the grading, the graders had access to the initial marks evaluated automatically by the system which consisted solely of the final numerical and multiple choice answers. The focus of the graders was to check if the handwritten solutions justified these answers and to adjust the preliminary marks accordingly. 

When there were grading disagreements or the marking scheme was deemed inadequate, graders had the option of contacting the AC via email. Discussion sessions over Zoom were also arranged for the graders. During the grading process, two errors in the official solutions of the theory round were discovered. Furthermore, some students came up with an alternative solution, which had not been included in the official marking scheme. Both situations were addressed by editing the marking scheme and notifying the graders about the changes.

Once all the grader pairs finished the initial round of grading, a spreadsheet containing marking discrepancies of all pairs was generated. In case of minor discrepancies (up to 0.5 marks), the higher of the two scores was automatically accepted, while for larger discrepancies, the cases were flagged for a review by moderators. For consistency, the policy was to allocate each problem to a single moderator as far as possible. During this stage  of the grading procedure, it was discovered that some graders had not been able to find the handwritten solutions. These cases had to be re-evaluated completely by the AC members.

Lastly, the students were given a chance to appeal against the marks allocated. All appeals were regraded independently by moderators.

\section*{Implementation}
\label{sec:Implementation}
From the outset it was clear that no existing commercial web-based solution would be suitable for a competition like GeCAA. Based on the design principles stated above, a tailor-made package was developed, which was adapted from the environment viktoriinid.ee. This environment was originally designed for holding national level online quizzes in Estonia and has been used since 2017. Presently, it hosts more than 11 national online quizzes every year and the number has increased almost every year. Due to the pandemic in 2020, it was tested for use in the Olympiad environment. The Estonian national round for the  astronomy student-contest was the first one in Estonia to use this environment as a means of conducting the national Olympiad online, while also enabling foreign nationals to participate. In total, more than 600 students from approximately 20 countries participated in the 2020 event, compared to just about 40 Estonian students in previous years.

For the GeCAA, a modified version of viktoriinid.ee environment\footnote{\url{https://viktoriinid.ee}} was developed. The engine is written mostly in Javascript and runs on AWS (Amazon Web Service) server-less solution, which is designed to auto-scale itself based on the number of active users, using the Amazon Lambda framework. The engine is able to keep a session running, even if the user faces  disconnection for a moderate period, based on active session tracking. In case the system is unable to reconnect a live session, it creates a secondary session with the same user credentials, which can be merged with the previous session’s activity. All students' results are stored in JSON format as independent entities in a NoSQL database. All results are stored in the AWS S3 storage space. To make the competition more compatible with automatic marking, all tasks were split into sub-tasks, such that they could be answered with a single input. The system is able to automatically grade numeric, \emph{lingua franca} (in case a single language is acceptable as an answer, eg Latin names for stars) textual answers, including spelling mistakes, and different checkbox-type questions. For GeCAA, a new feature was developed in the system, which was to allow students to mark a location in a given image. The software was pre-fed an accepted image region for each correct answer, allowing the student’s marking to be graded automatically. This feature was useful in the questions related to the sky map and sky images, where the students were asked to mark astronomical objects in their correct positions on the image.

The environment did not create a new interface for submitting questions and translations. Instead, it imported questions and translations from {\emph{Google Docs}} at the start of the examination session. This allowed an easy input mechanism for original versions of the questions and easy collaboration for translations. Even for the student interface, the students could toggle between any translations with a simple click of a button.

Alongside digital inputs, students were asked to submit their handwritten work through the competition environment. In case of any difficulties, as a secondary backup solution, they were given an option to upload their work on pCloud cloud disk or to submit it via e-mail, from where an automated script was used to transfer them, first to Google Drive and then into pCloud. 

The environment was opened two weeks before the actual event, for students and team leaders to test features and get familiarised with the interface. Students were also given a chance to take some mock tests.

The grading process for all the tasks involved each solution being independently graded by two graders. The system compared the scores of the two graders and flagged up the discrepancies, if any, for a head examiner of each task to resolve. Post resolution, the system was able to do auto-totalling and generate a rank order.

Some lessons learned from the software implementation were:
\begin{itemize}
\item On the software side, given the fact that the system could only check for numerical values or tick boxes and textual values which followed a specific pattern (e.g., Latin names for constellations), it is important to design questions which would include final answers in this form. Any diagrams or plots included in scanned worksheets could not be graded electronically. The software system for uploads should be tested rigorously by the participating teams during the trial phase. During GeCAA, several teams faced last minute difficulties in including scanned sheets in the software system and chose to upload them to pCloud later. Matching these pCloud-based images to the students who submitted them turned out to be difficult and time-consuming for the LAC and created confusion for the graders.
\item The software system used for GeCAA was adapted from an existing system used in Estonia. Nonetheless, a number of new features were added and the eventual load on the system was higher than anything previously tested. This required a support team to provide real-time solutions to glitches, including working on weekends. This means that even an online competition based on existing software needs to budget for an expert IT support team.

\end{itemize}

\section*{The National Astronomy Competition for the USA}
The National Astronomy Competition (NAC), an annual event, serves as the selection process for the US team for the International Olympiad on Astronomy and Astrophysics (IOAA). The selection is run by the United States Astronomy and Astrophysics Olympiad (USAAAO), a volunteer-run non-profit organization, and includes two proctored rounds: a multiple-choice test in February, and a free-response exam in March or April, both covering the entire IOAA syllabus. Registration for these exams typically begins in early December and can be completed at the USAAAO website (https://usaaao.org/). The first round requires a \$25 registration fee, but case-by-case financial aid is available. High-scoring students from the first round are invited to the NAC. The top ten performers at the NAC are then invited to represent the USA at the IOAA, for which the costs are fully covered by USAAAO, with additional students being invited to participate in several months of free online training based on their performance and other factors.

\section*{International Competitions}

Competitions in science at high school and college levels, such as the International Physics Olympiad \citep[IPhO,][]{Khoi2009,MoranLopez2010,Kusamran2012, Kalda2013a}, the
International Young Physicists' Tournament  \citep[IYPT,][]{Planinsic2009,Binder2009,Rajkovits2010,Kunesch2010,Kewei2011,Kim2012,Chan2014,Ren2015,Plesch2017,Landgraf2017,Plesch2018,Wen2020,Plesch2020,Liu2021,Wen2021, Jaikumar2022}, the International Physicists' Tournament \citep[IPT,][]{Vanovskiy2014,Forro2014,Michalke2020}, the Physics League Across Numerous Countries for Kick-ass Students \citep[PLANCKS,][]{Haasnoot2016, Dorn2018, Heeboll2020, Rini2021}, the World Physics Olympiad \citep[WoPhO,][]{Kwee2013}, the European Physics Olympiad \citep{Heidelberg2018}, the ``Physics Cup'' \citep{Kalda2013}, First Step to Nobel Prize in Physics competition \citep{Gorzkowski2011}, the Slovene Science competition \citep{Rovsek2017} part of the the Chain Experiment \citep{Dziob2017,Susman2017}, the German Physics Olympiad \citep{Petersen2017} and
the German Physicists' Tournament \citep[GPT,][]{Bley2021} have long been shown to enrich and support students scientific development \citep{Sahin2015,Stake2001}. IOAA fills this role as well, focusing on astronomy and astrophysics education.

\section*{Declarations}

\paragraph{List of abbreviations}
\begin{itemize}
\item AC - (International) Academic Committee
\item AGNs - Active Galactic Nuclei
\item AWS - Amazon Web Service
\item GeCAA - Global e-Competition on Astronomy and Astrophysics
\item IOAA - International Olympiad on Astronomy and Astrophysics
\item IPhO - International Physics Olympiad
\item LAC - Local (Estonian) Academic Committee
\end{itemize}

\paragraph{Ethical Approval}
The present work does not include any data collection of human subjects focused on behaviour or understanding of any educational concept. The only survey mentioned in the manuscript is anonymous non-mandatory feedback, submitted voluntarily, about a public event, and hence is exempt from any ethical approvals.


\end{document}